\documentclass[aps,pra,twocolumn,showpacs]{revtex4}

\usepackage{epsfig}
\usepackage{graphicx}
\usepackage{dcolumn}
\begin{document}
\title{Optimizing the fast Rydberg quantum gate}
\author{M. S. Safronova}
\author{Carl J. Williams}
\author{Charles W. Clark}
 \affiliation{
Physics Laboratory,
National Institute of Standards and Technology,
Technology Administration,
U.S. Department of Commerce, 
Gaithersburg, Maryland 20899-8410}

\date{\today} 

\begin{abstract}
The fast phase gate scheme, in which the qubits are 
atoms confined in sites 
of an optical lattice, and gate operations are 
mediated by excitation of Rydberg states, was proposed by
Jaksch {\em et al.} Phys.\ Rev.\ Lett.\ {\bf 85}, 2208 (2000).
A potential source of decoherence
in this system derives from motional heating,
which occurs if the ground and Rydberg states
of the atom move in different optical lattice potentials.
We propose to minimize this effect by choosing the lattice
photon frequency $\omega$ so that the ground and Rydberg states
have the same frequency-dependent polarizability
$\alpha ( \omega )$. The results are presented for the case of Rb.   
\end{abstract}


\maketitle

    Recently, a number of schemes for quantum computation with 
    neutral atoms have been proposed \cite{qp1,qp2,qp3,qpn1,qp4,qp5,qpn2,1,qpn3}. 
In those schemes, qubits are realized as internal states of     
     neutral atoms trapped in optical
     lattices or magnetic microtraps. 
     The two-qubit quantum gates are realized using
    controlled cold collisions
\cite{qp1,qp2,qp3,qpn1}, 
 controlled dipole-dipole 
interactions \cite{qp4,qp5,qpn2}, or       
   by conditional excitations of atoms into the Rydberg states
by a series of laser pulses \cite{1,qpn3}. This approach to quantum 
computation has many advantages, such as scalability, possible 
massive parallelism, long 
decoherence times of the internal states of the atoms, 
flexibility in controlling atomic interactions, 
and well-developed experimental techniques. As extensive
experimental studies of the feasibility of using neutral atoms 
for quantum computation are under way, more detailed theoretical 
investigation of the various schemes is needed. In this work, we 
investigate optimization of the quantum gate scheme realized by  
excitations to Rydberg states \cite{1}. We refer to this scheme 
below as the Rydberg gate. The choice of this particular scheme 
results from its potential for fast (sub-microsecond) gate operations. 

In the Rydberg gate scheme, the basic qubit is based on two ground 
hyperfine states of neutral atoms confined in an optical lattice.  
A two-qubit phase gate may 
be realized by conditionally exciting two atoms to low-lying 
Rydberg states. Different versions of the scheme have been proposed \cite{1}. 
In those schemes, either one or both atoms may occupy the 
Rydberg state for much of the duration of the gate operation. 
However, an atom 
in a Rydberg state will, in general, move in different optical 
lattice potential than that experienced by the ground state.  
Therefore, the 
vibrational state of the atom in the lattice
may change after the gate operation is completed,
leading to decoherence due to motional heating.
The optical potential for a given state depends on its ac
polarizability, so we can seek to minimize this
motional heating effect by the choice of a 
particular Rydberg state or of the lattice photon frequency, $\omega$. 
In this paper, we describe a method for
accomplishing this by
matching the frequency-dependent polarizabilities, $\alpha(\omega)$, of the
atomic
ground state and Rydberg state. 
The results are presented for case of Rb; however,
the approach used here is applicable for logic gates with other 
alkali-metal atoms.

In this work, we calculate polarizabilities of the Rb atom 
in its ground state, $\alpha_{5s}(\omega)$, and in various 
   Rydberg $ns$ states $\alpha_{ns}(\omega)$.
    We demonstrate that there exists 
   a lattice frequency  $\omega$ such that 
   \begin{equation}
   \alpha_{5s}(\omega)\approx \alpha_{ns}(\omega).
   \end{equation}
This finding relies upon the fact \cite{Claude} that the 
frequency-dependent polarizabilities of
alkali Rydberg states are close to that of a free electron,
$\alpha_{\rm free} (\omega ) = - e^2/m_{\rm e} \omega^2$,
where $e$ is the elementary charge in statcoulombs, and 
$m_{\rm e}$ is the electron mass.  For the optical
frequencies treated in this paper, 
$\alpha_{nl}(\omega)$ for Rydberg states is comparable 
in magnitude to $\alpha_{5s}(0)$, 
but it has a negative sign.  As a function
of $\omega$, $\alpha_{5s}(\omega)$
increases as $\omega$ increases from zero, but changes sign
when $\omega$ exceeds the first resonance frequency.  We show
that it will always be possible to find a frequency for which
the Rydberg and ground-state polarizabilities are equal.  An 
optical lattice constructed with light of that frequency will
have well depths characteristic of far-off-resonance traps. 

An alternative approach to polarizability matching of ground
and Rydberg states is to find Rydberg states with resonant
frequencies close to those of the ground state.  In such
cases the polarizabilities can be matched at near-resonant frequencies,
resulting in a tighter trap for a given laser
intensity. Such possibilities can be identified
by a straightforward search of tabulated atomic energy levels,
and they turn out to be rare.  The two cases of interest
in Rb are discussed at the end of this paper.

We begin with a description of our calculation of the 
polarizabilities. It is convenient for this purpose to use the 
system of atomic units, a.u., in which $e, m_{\rm e}$, $4\pi 
\epsilon_0$ and the reduced Planck constant $\hbar$ have the 
numerical value 1.  Polarizability in a.u. has the dimensions of 
volume, and its numerical values presented here are thus measured 
in units of $a^3_0$, where $a_0\approx0.052918$~nm is Bohr radius.
The atomic units for $\alpha$ can be be converted to SI units via
 $\alpha/h$~[Hz/(V/m)$^2$]=2.48832$\times10^{-8}\alpha$~[a.u.], where
 the conversion coefficient is $4\pi \epsilon_0 a^3_0/h$ and 
 Planck constant $h$ is factored out. The atomic unit of frequency $\omega$
 is $E_h/\hbar\approx4.1341\times10^{16}$~Hz, where $E_h$ is Hartree
 energy. 
 
  The 
valence contribution to the dynamic polarizability for an alkali 
atom in an $ns$ state can be calculated using the formula (in 
a.u.) \cite{relsd} 
\begin{eqnarray}
\alpha_{ns}(\omega)&=&\frac{1}{3}\sum_{n^{\prime}}
\left(
\frac{(E_{n^{\prime}p_{1/2}}-E_{ns})
\langle n^{\prime}p_{1/2}\|D\| ns \rangle^2 }
{(E_{n^{\prime}p_{1/2}}-E_{ns})^2-\omega^2}\right. \nonumber  \\
&+&\left. \frac{(E_{n^{\prime}p_{3/2}}-E_{ns})
\langle n^{\prime}p_{3/2}\|D\| ns \rangle^2 }
{(E_{n^{\prime}p_{3/2}}-E_{ns})^2-\omega^2}
\right),
\label{eq1}
\end{eqnarray}
where $D$ is the electric dipole operator and $E_{i}$ is the energy of the
state $i$.
In this formula,  $\omega$ is 
assumed to be at least several linewidths off resonance with the 
corresponding transition. This condition is satisfied for the 
frequencies considered in this work, as the lattice has to be significantly 
detuned from the resonance to minimize spontaneous emission.    
The core contribution to the polarizability, calculated in 
the Hartree-Fock (HF) approximation, is 
found to be small (9.3~$a_0^3$) and is weakly dependent on $\omega$ 
in the frequency range considered here.  
The static value for the polarizability
of Rb$^+$ 
calculated in the random phase
approximation (RPA) \cite{der,jkh} is 9.1 $a_0^3$,
close to the value of 9.0 $a_0^3$ obtained from
by Johansson \cite{joh} from analysis of the 
observed term values
of nonpenetrating Rydberg states.
The accuracy of the core contribution is estimated to be 5~\%
in Ref.~\cite{relsd}.
We use the RPA value for the core polarizability
of Rb$^+$ as a baseline, and adjust it
to account for valence electron and
the frequency
dependence by using HF calculations. 
The RPA and HF values differ by only 2~\%.

In calculating the valence contribution,
the term with $n=5$ dominates the the sum over states,
and the contributions from higher values
of $n$ converge rapidly for all values of $\omega$
considered in this work.  
 The $5s-5p$ matrix elements
 were measured in Ref.~\cite{2} and 
 the $5s-np$ matrix elements with $n=6, 7, 8$ were calculated 
 using single-double all-order method in Refs.~\cite{relsd,4}.
 We have conducted the calculation of the ground ($5s$) state
 $\alpha_{5s}(\omega)$ in Rb, including terms with $n=5, 6, 7$, and 8
 using matrix elements from Refs.~\cite{2,4} and experimental
 energies from \cite{exp}.
  The contributions to the ground state polarizability from states with 
 $n^{\prime}=8-\infty$, including the continuum, and 
 from terms with $n=2, 3, 4$ are estimated to be very
  small (0.2~$a^3_0$ and -0.3~$a^3_0$, respectively). 
  The results for the values of $\omega$ near first two resonances, 
  $5s-5p_{1/2}$ and $5s-5p_{3/2}$ are listed in Table~\ref{tab1n}.     
     As expected, the value of the frequency-dependent polarizability varies
very   
   rapidly in the vicinity of the $5s-5p_{1/2}$  and $5s-5p_{3/2}$
   resonances. The second set of values of $\alpha_{5s}(\omega)$ listed in 
    column four illustrates the existence of the $\alpha_{5s}(\omega)=0$
    point between the $5s-5p_{1/2}$ and $5s-5p_{3/2}$ resonances.  
    The next resonance occurs at  $\omega=E_{5s}-E_{6p_{1/2}}$.  
  
  \begin{table}
\caption{\label{tab1n} Dynamic polarizabilities $\alpha(\omega)$ 
of the ground state of Rb in units of $a_0^3$;  $\omega$ is in 
a.u. ($E_h/\hbar$).} 
\begin{center}
\begin{ruledtabular}
\begin{tabular}{lrrrrrlr}
\multicolumn{1}{c}{$\omega$}&
\multicolumn{1}{c}{$\alpha$}&
\multicolumn{1}{c}{}&
\multicolumn{1}{c}{$\omega$}&
\multicolumn{1}{c}{$\alpha$}&
\multicolumn{1}{c}{}&
\multicolumn{1}{c}{$\omega$}&
\multicolumn{1}{c}{$\alpha$}\\
   \hline
 0       &  318.5(6)&&   0.0576500& -815 (23)&& 0.0578 &     3931(20)   \\
 0.02000 &  360.0(7)&&   0.0576645& -290 (23)&& 0.058 &    10755(31)\\
 0.04000 &  597.8(8)&&   0.0576700&  -97 (22)&& 0.059 &   -11548(16)\\
 0.04298 &  693.5(9)&&   0.0576722&  -21 (22)&& 0.060 &    -4737(5)\\
 0.05000 &   1211(1)&&   0.0576728&    0 (22)&& 0.065 &    -1206(1)\\
 0.05500 &   3132(3)&&   0.0576734&   21 (22)&& 0.070 &    -667.1(9)\\
 0.05700 & 13854(23)&&   0.0576800&  246 (22)&& 0.080 &    -330.3(8)\\
 0.05750 & -9311(57)&&   0.0576900&  581 (22)&& 0.090 &    -205.9(8)\\
 0.05760 & -2869(29)&&   0.0577000&  907 (21)&& 0.100 &    -138(1)\\
\end{tabular}
\end{ruledtabular}
\end{center}
\end{table}
     
Molof et al. \cite{Molof} measured $\alpha_{5s}(0)$
to be $47.3 \pm 0.9~$\AA$^3$, equivalent
to $319 \pm 6~a_0^3$.  Our calculated value agrees with this result to 
within the experimental uncertainty. The static polarizability of the ground state
 of Rb, $\alpha_{5s}(0)$, was previously calculated using the techniques
 described above
in Ref.~\cite{der}. 
The general behavior of the frequency-dependent
polarizability of the ground state of 
Rb has been investigated previously in \cite{1a}. A result for one
    particular value of $\omega=0.04298$~a.u., corresponding to 
$\lambda=1.06 \mu$m \cite{wavelength_comment},
    has been calculated in the same work, using
a model potential method. We compare our 
  value of the ground state 
dynamic polarizability for this frequency with the result of 
Marinescu {\it et al.} \cite{1a}. Our result  
$\alpha=693.5(9)~a_0^3$, listed in Table~\ref{tab1n}, 
 is in good agreement with the result 
$\alpha=711.4~a_0^3$ from Ref.~\cite{1a}. 
  It is just outside the range of uncertainty
of the value $\alpha=769 \pm 61~a_0^3$ inferred by
Bonin and Kadar-Kallen from an atomic deflection 
experiment \cite{Bonin}. 
 A subsequent publication \cite{tbs} will provide
further details of the polarizability calculations, as well as 
 calculation of the radiative widths of $np$ levels to 
 quantify the behavior of the ground state polarizability near 
 the resonances. 

 Second, we calculate the dynamic polarizabilities of the Rydberg states
for a specific value of the lattice photon frequency $\omega$ 
and then investigate the dependence of polarizability on $\omega$ 
for several Rydberg states.   The most likely realization of the 
Rydberg gate scheme involves two-photon transitions from the 
ground state to either $ns$ or $nd$ states. For clarity, we 
calculate the polarizabilities of the $ns$ states in this work.  
 To make an estimate of the $ns$ Rydberg state polarizability
$\alpha_{ns}(\omega)$ we 
first calculate relevant matrix elements and  energies
 in Hartree-Fock approximation. 
 The HF calculations are done on a non-linear grid, and the question of the
numerical stability 
 of the calculation of the 
 properties of the Rydberg states was investigated by conducting the 
 calculation with several different grids to insure the stability of the 
 results. 
We use the resulting matrix elements to calculate frequency-dependent
polarizabilities 
 of the $ns$ states in HF approximation. 
 The results for the $ns$ state polarizabilities
 with  $\omega=0.0576645$~a.u.
  are listed in Table~\ref{tab3} in the columns labeled
$\alpha_{\mathrm{HF}}$.
  This frequency corresponds 
 to the value of the $\alpha_{5s}(\omega) = -290~a_0^3$ 
 and to a detuning approximately 1/3 of the distance between the 
 two resonances; $\alpha_{\rm free} (\omega ) = -301~a_0^3$ at
this frequency.

\begin{table}
\caption{\label{tab3} Dynamic polarizabilities $\alpha_{ns}(\omega)$ (in units of $a_0^3$)
for Rb, $\omega$=0.0576645~a.u.,
i.e. $\lambda = 790$ nm. } 
\begin{center}
 \begin{ruledtabular}
 \begin{tabular}{cccccc}
\multicolumn{1}{c}{$N$}&
\multicolumn{1}{c}{$\alpha_{\mathrm{HF}}$}&
\multicolumn{1}{c}{$\alpha$}& \multicolumn{1}{c}{}& 
\multicolumn{1}{c}{$N$}& 
\multicolumn{1}{c}{$\alpha_{\mathrm{HF}}$}\\ 
    \hline
    8   &     -304      &  -295\footnotemark[1]&&  14    & -286  \\
    9   &     -292      &      &&  15    & -285  \\
   10   &     -289      &      &&  16    & -284  \\
   11   &     -288      &      &&  17    & -282  \\
   12   &     -287      &      &&  18    & -280  \\
   13   &     -287      &      &&  19    & -277  \\
\end{tabular}
\end{ruledtabular}
\end{center}
\footnotetext[1] {\noindent High-accuracy value obtained  using 
  experimental energies and  all-order matrix 
  elements for the dominant terms with $n^{\prime}=7,~8$.}
\end{table} 
  
  We observe that there is no significant change of the 
  polarizability values after $n\geq 8$.
    The summation over $n$  in 
  Eq.~(\ref{eq1})  is  truncated at $n=23$.  The contribution of the 
  states with $n>23$ and continuum is evaluated by carrying out a
calculation 
  of the polarizability of the $10s$ state with B-spline basis set
\cite{wrj}.
  The summation over the entire basis set yields the result $-292$~a.u.
   which differs from the value in Table~\ref{tab3} 
  by only 1~\%.  To evaluate the uncertainty of the 
  HF approximation further we repeat the calculations using HF matrix
elements and 
 experimental energies from Ref.~\cite{exp} in Eq.~(\ref{eq1}). 
We find a substantial, from 10~\% to 30~\%, difference
 between these two approximations despite only a few percent differences
between
 Hartree-Fock and experimental $[E_{ns}-E_{np}]$ energies. Such a large  
 discrepancy is explained by severe cancellations of the different terms  in
the sum 
 of Eq.~(\ref{eq1}). 
 As an illustration, we list the contributions to the polarizability of the
$15s$ state from 
 several dominant  terms together with the values for the 
 corresponding  dipole matrix elements $\langle np\|D\|15s\rangle$, 
 energy differences $\delta E=E_{np}-E_{15s}$, and denominators  of 
 Eq.~(\ref{eq1}) in   
 Table~\ref{tabcontr}. We find that the dominant contributions 
 come from $n=14$ and $n=15$ terms which have different sign owing 
 to the sign change in the energy differences.  We also note
 that the denominator $(\delta E)^2-\omega^2$ is completely dominated by the
 $\omega^2$ term, and, therefore, nearly identical for all of the
contributions.
     We also note the cancellations of the smaller pairs of terms with
$n=13,~16$
     and $n=12,~17$.

 \begin{table}
\caption{\label{tabcontr} Contributions to the dynamic 
polarizability of the $15s$ state of Rb in a.u., 
$\omega$=0.0576645~a.u., $\delta E=E_{np}-E_{15s}$, 
$D=\langle np\|D\| 15s \rangle$. The contributions from different 
terms are given in column $\alpha(\mathrm{contr})$. The 
accumulated values are given in column $\alpha(\mathrm{acc})$. } 
\begin{center}
 \begin{ruledtabular}
 \begin{tabular}{crrrrr}
\multicolumn{1}{c}{$np$}& \multicolumn{1}{c}{$D$}& 
\multicolumn{1}{c}{$\delta E$}& \multicolumn{1}{c}{$(\delta 
E)^2-\omega^2$}& \multicolumn{1}{c}{$\alpha(\mathrm{contr})$}& 
\multicolumn{1}{c}{$\alpha(\mathrm{acc})$}\\ 
    \hline
     $12p_{1/2}$ &     -5.9  &  -0.00215 &  -0.00332&     8 &     28\\
     $12p_{3/2}$ &     -8.2  &  -0.00214 &  -0.00332&    14 &     42\\
     $13p_{1/2}$ &    -15.6  &  -0.00111 &  -0.00332&    27 &     69\\
     $13p_{3/2}$ &    -21.7  &  -0.00110 &  -0.00332&    52 &    122\\
     $14p_{1/2}$ &   -111.8  &  -0.00034 &  -0.00333&   424 &    545\\
     $14p_{3/2}$ &   -161.0  &  -0.00033 &  -0.00333&   858 &    1403\\
     $15p_{1/2}$ &    144.0  &   0.00026 &  -0.00333&  -536 &     867\\
     $15p_{3/2}$ &    201.2  &   0.00026 &  -0.00333&  -1072&    -204\\
     $16p_{1/2}$ &     16.1  &   0.00073 &  -0.00332&  -19 &     -223\\
     $16p_{3/2}$ &     23.9  &   0.00073 &  -0.00332&  -42 &     -265\\
     $17p_{1/2}$ &  6.4  &   0.00110 &  -0.00332&   -5 &     -270\\
     $17p_{3/2}$ &  9.7  &   0.00110 &  -0.00332&  -10 &     -280\\
  \end{tabular}
\end{ruledtabular}
\end{center}
\end{table}

                         \begin{figure}  [ht]
\includegraphics[scale=0.45]{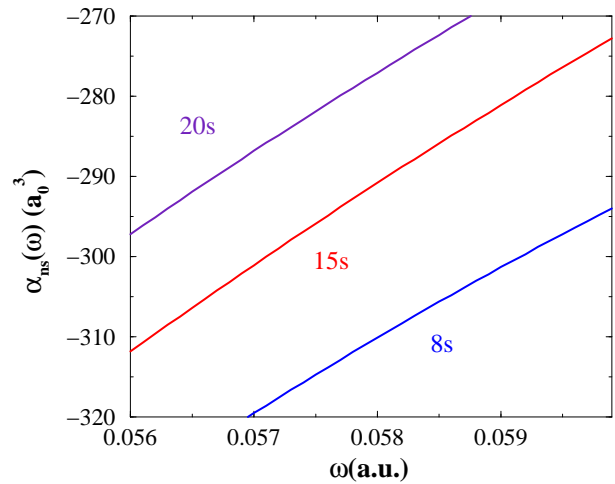}   
\caption{ \label{fig4} Dynamic polarizability $\alpha(\omega)$ for 
the $ns$ states  of Rb in atomic units.  }             
\end{figure} 
                 
 The  cancellations of dominant contributions have also been 
 observed in the calculations of the vector transition 
 polarizability $\beta$ and parity-nonconserving amplitude in Cs and Fr (
 see, for example \cite{relsd,4} and references therein) when calculated 
 using similar ``direct-summation'' method.
  Generally, the  values obtained using sets of data of the 
  consistent accuracy are more reliable 
  in such cases, so we use HF data below.     
The accuracy of the 
  values can be increased by estimating correlation correction
  contributions to the values of the dipole matrix elements.  
  We conduct such calculation for $\alpha_{8s}(\omega)$
  for the same frequency as data in Table~\ref{tab3} using 
  experimental energies and high-accuracy all-order matrix 
  elements for the dominant terms with $n^{\prime}=7,~8$. The result (-295~$a^3_0$)
  is listed in Table~\ref{tab3} and agrees very well with 
  Hartree-Fock value. We note that substituting HF energy by 
  experimental values and leaving matrix elements unchanged yields 
  substantially different (-356~$a^3_0$) value, thus confirming our 
  conclusion that using HF values for both matrix elements and 
  energies produces more accurate values then replacing HF 
  energies by experimental results. 
   We note that current accuracy of $\alpha_{ns}(\omega)$ (estimated at
10~\%) is sufficient for the 
 purpose of present paper since the polarizability of the ground state 
 varies vary rapidly with $\omega$ as illustrated in Table~\ref{tab1n}
 and the variation of the value $\omega$ within the uncertainty of the 
 Rydberg state polarizability is small. Consequently, the  exact point
 where  $\alpha_{5s}(\omega) = \alpha_{ns}(\omega)$ can be      
 determined experimentally by detuning of the lattice frequency near the 
 matching point.  Further investigations of the 
 Rydberg state polarizabilities are beyond the scope of this paper and will 
 be considered elsewhere \cite{tbs}.

  Finally, we  investigate the dependence of the 
Rydberg state polarizabilities on the value of $\omega$ in the 
vicinity of the $5s-5p_{1/2}$ and $5s-5p_{3/2}$ resonances. As 
noted above, the polarizability of $ns$ state with $n>8$ does not
change significantly with $n$. The results of the calculation of 
valence dynamic polarizability for the $8s$, $15s$, and $20s$ 
states with $\omega=0.056-0.060$~a.u. 
 are illustrated  in Fig.~\ref{fig4}. We find that 
 the dynamic polarizabilities of the Rydberg $ns$ states vary 
 very weakly with $\omega$.  
          
    \begin{figure} 
 \hbox{\centerline{\epsfxsize=3.3in\epsffile{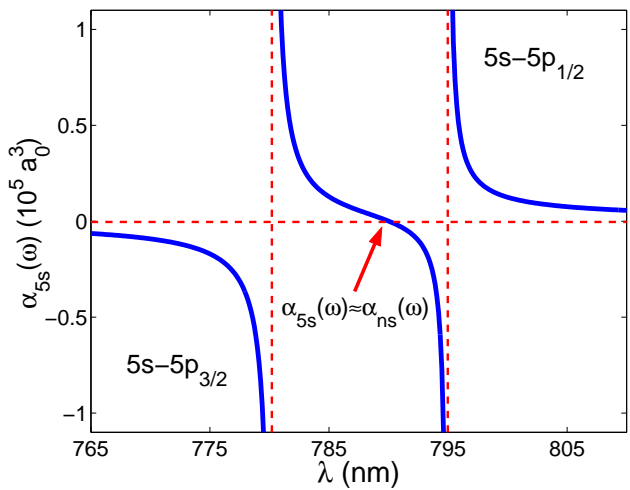}}}
   \caption{Dynamic polarizability $\alpha(\omega)$ for the ground state 
   of Rb in a.u.  The horizontal 
dashed line  corresponds to polarizability of the Rydberg $ns$ 
level. } 
    \label{fig5}     
   \end{figure}      

 Summarizing the results above we find that the values of the 
 ground state and Rydberg $ns$ state polarizabilities can be 
 matched near the point between $5s-5p_{1/2}$ and 
 $5s-5p_{3/2}$ resonances where the ground state polarizability 
 changes sign as demonstrated in Fig.~\ref{fig5}.
 The 
horizontal dashed line  corresponds to polarizability of the 
Rydberg $ns$ level.
 The resonances are shown by the vertical dashed lines for clarity as we 
   assumed $\omega$ to be a few linewidths from the resonances.
  We predict that 
 value of $\omega$ where these polarizabilities match equals to
  $\lambda_{vac}=790.14(2)$~nm for $15s$ state.
  This value includes 10~\% uncertainty in the value of the 
  15s state polarizability and 6~\% uncertainty in the value of the 
  ground state polarizability. 
  The next point at which the polarizabilities
     can be matched is close to the $5s-6p_{1/2}$ resonance which requires
     much larger detuning.   
The above discussion is valid for higher Rydberg states, which may 
be chosen for the  gate implementation because of the longer
lifetimes \cite{M},  
since we found Rydberg state ac polarizability to vary weakly with the 
principal quantum number.

\begin{figure} 
\includegraphics[scale=0.9]{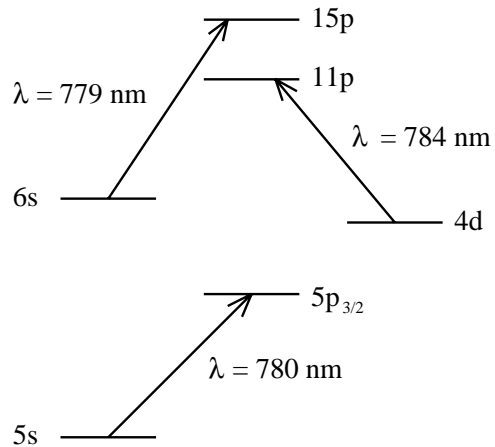}   
\caption{ \label{fig2} Levels with transitions energies near 
$5s-5p_{3/2}$ resonance. Levels are not to scale.}             
\end{figure}

 Next, we consider an alternative approach to the issue of the 
 matching polarizabilities of the ground and arbitrary Rydberg state.
 This discussion is not limited to $ns$ Rydberg states.   
 It follows from the expression for the dynamic polarizability 
 that if the energy 
 difference $E_{ns}-E_{n^{\prime}p}$ or $E_{nd}-E_{n^{\prime}p}$ 
  is accidentally close to the value of 
 either $5s-5p_{1/2}$ or $5s-5p_{3/2}$  transition energy then 
 the polarizability of the corresponding $s$, $p$, or $d$ state
 can be made large by detuning to the appropriate frequency. 
  We have investigated the spectrum 
     of Rb to locate energy differences which are close to the resonance   
  $5s-5p_{1/2}$ or    $5s-5p_{3/2}$  transition energies. 
  The closest matches are 
     $6s-15p$ and $4d-11p$ transition energies  which
     differ from the $\delta E=E_{5s}-E_{5p_{3/2}}$ by $20$~cm$^{-1}$ and
$60$~cm$^{-1}$, 
     respectively. 
The corresponding energy level scheme is illustrated in Fig.~\ref{fig2}. 
      However, in both of these cases the terms with such denominators
      contribute to the polarizabilities of the $p$ states.  
       The required detuning in this case is smaller
 but only $11p$ and $15p$ levels may be used in this method.     
      We note that there are no such Na levels for the
       situation where the lattice
      light is near Na(3p) levels.    
      The advantage of this scheme is high value of the ac 
     polarizabilities at the matching point; in the previous scheme 
     the value of the ac polarizability at the matching point is 
     relatively small, leading to the higher laser power requirement and 
     subsequent higher scattering rate (on the order of 10~$s^{-1}$).  
    The disadvantage of using $np$ vs. $ns$ states in a Rydberg
    gate scheme is that excitation of transitions from the ground
state requires either ultraviolet radiation 
or a three-photon process. Another difficulty of this approach 
 is the necessity of the 
very fast gate operation times to achieve high gate fidelity
 because of the short, 4~$\mu s$ \cite{T}, lifetime 
of the $15p$ level.   

 We note that by choosing the appropriate Rydberg level and 
 a longer wavelength trapping laser (e.d. CO$_2$ laser) 
 it should often be possible to find a matching scheme. 
 In these situations, because of the number of nearby resonances 
 for the Rydberg state, one probably would have to consider contributions from
 several intermediate states. 
        
In summary, we have found two ways to match frequency-dependent
polarizabilities 
of the Rydberg state and the ground state for optical lattices
tuned near the Rb(5p) states. First, the value of the 
ground state polarizability at
 $\lambda_{vac}=790.032(8)$~nm crosses zero and changes sign. Therefore, it
can 
be matched with polarizabilities of various Rydberg levels near 
that point as demonstrated in Fig.~\ref{fig5}. 
 For the case of $15s$ state, polarizabilities match at 
$\lambda_{vac}=790.14(2)$~nm. 
  Second, the value of $\omega$ can be 
detuned to nearly match the values of the $15p-6s$ and $11p-4d$ 
transition energies  
 which are accidentally close to the $\delta E = E(5p_{3/2})-E(5s)$ energy.
 Then, the dynamic polarizability of the $15p$ or $11p$ state can be made
large
 enough to match the polarizability of the ground state near
 $5p_{3/2}-5s$  resonance. 
 The matching of the frequency-dependent polarizabilities of the atom in its
ground and 
 Rydberg states results in the matching of the optical potential
 seen by the atom during the gate operation and
  provides the optimal scheme for the Rydberg gate 
 operation  with respect to the motional decoherence.

Another way of eliminating the differential in the trapping potential
between the ground and Rydberg states is to switch the trap off during
the time of the gate action, and turn it back on after the action is
completed.  This procedure also induces heating of the
atomic center of mass motion, the effect of which can be 
characterized as follows.  Suppose that, at time $t=0$, 
an atom of mass $M$
is in the ground state of a
trap site, where for convenience we assume the potential to be that
of an isotropic oscillator of frequency $\omega_0$,
with ground state energy $E_0 = 3 \hbar \omega_0/2$.  
If the trap is turned
off suddenly, the wavefunction  
$\Psi\{{\mathbf{r}},t\}$ for the center of mass 
coordinate $\mathbf{r}$
evolves during time t as

\begin{equation}
\Psi\{{\mathbf{r}},t\} = \frac{\exp(-r^2/2d_0^2 (1 + i \omega_0 t))}
{\pi^{3/4} d_0^{3/2} (1 + i \omega_0 t)^{3/2}},
\end{equation} 
where $d_0 = \sqrt{\hbar/M \omega_0}$.  In free expansion
of the atomic wavepacket from the released trap, its mean
kinetic energy, $<T> = E_0/2$ remains constant,
while its  mean-square 
radius grows as $<r^2(t)> = 3 d_0^2 (1 + \omega_0^2 t^2)/2$.  If the
trap is suddenly turned back on again at time $\tau$, the mean
energy of the wavepacket in the restored trapping potential 
is $E = E_0/2 + M \omega_0^2 <r^2(\tau)>/2 = 
E_0 (1 + \omega_0^2 \tau^2/2)$.
This corresponds to a heating of 
$k_{\rm B}T = (\hbar \omega_0)  \omega_0^2 \tau^2/4$ per cycle
of trap release and restoration.  Such heating can be avoided
by use of a matched potential scheme of the type proposed here.
For a trap frequency of a 1~MHz and a 1~$\mu$s gate time one 
obtains $k_B T\approx 0.006(\hbar \omega_0)$.
 The gate operation time in principle can be optimized by appropriate 
 choice of the Rydberg level, the applied dc field, and laser power.
 
We gratefully acknowledge discussions with Steven Rolston, Wendell 
Hill, Walter Johnson, Philippe Grangier, Mark Saffman, and  Thad Walker.
 This work was partially supported by the Advanced
Research Development Activity, the National Security Agency, 
and NIST Advanced Technology Program.


\begin{thebibliography} {}
\bibitem{wavelength_comment} We use the value
$\omega=0.04298$~a.u. that was used in Ref. \cite{1a}, but
we note that the value $\omega=0.04281$~a.u. actually corresponds to the
wavelength in air of $\lambda=1.064 \mu$ of the Nd:YAG laser.
\end{thebibliography}
\end{document}